\documentclass[11pt]{revtex4}
\pdfoutput=1
\usepackage{color,epsfig}
\usepackage{bm}
\usepackage{amsmath,amsfonts,amssymb,bm}
\usepackage{hyperref}
\usepackage{subfigure}
\newcommand{\pd}{\partial}
\newcommand{\bea}{\begin{eqnarray}}
\newcommand{\eea}{\end{eqnarray}}
\newcommand{\be}{\begin{equation}}
\newcommand{\ee}{\end{equation}}

\newcommand\bmb{\left( \begin{matrix}}
\newcommand\emb{\end{matrix} \right)}
\renewcommand{\(}{\left(}
\renewcommand{\)}{\right)}

\renewcommand\vec{\bm}

\begin{document}
\title{Electronic transport in torsional strained Weyl semimetals}

\author{Rodrigo Soto-Garrido}
\affiliation{Facultad de Ingenier\'ia y Tecnolog\'ia, Universidad San Sebasti\'an, Bellavista 7, Santiago 8420524, Chile} 

\author{Enrique Mu\~{n}oz}
%\email{munozt@fis.puc.cl}
\affiliation{Facultad de F\'isica, Pontificia Universidad Cat\'olica de Chile, Vicu\~{n}a Mackenna 4860, Santiago, Chile}

\date{\today}

\begin{abstract} In a recent paper \cite{Munoz-2017} we have studied the effects of mechanical strain and magnetic field on the electronic transport properties in graphene. In this article we extended our work to Weyl semimetals (WSM). We show that although the WSM are 3D materials, most of the analysis done for graphene (2D material) can be carried out. In concrete, we studied the electronic transport through a cylindrical region submitted to torsional strain and external magnetic field. We provide exact analytical expressions for the scattering cross section and the transmitted electronic current. In addition, we show the node-polarization effect on the current and propose a recipe to measure the torsion angle from transmission experiments.

\end{abstract}

\maketitle

\section{Introduction}
\label{sec:Intro}

At the dawn of relativistic quantum mechanics, H. Weyl discovered in 1929 \cite{Weyl_29} a solution to the Dirac equation in terms of massless particles, as an attempt
to model the physics of neutrinos. However, since neutrinos do have a finite mass, as has been confirmed
with experiments in high-energy physics \cite{Hirata_88}, they cannot be described as Weyl fermions. Nevertheless, the mathematical structure of these solutions, that in addition to spin possess an intrinsic
property called chirality (the projection of spin over the momentum direction), found much later a realistic physical scenario  in the context of condensed matter, as an effective low-energy model
for a class of materials that exhibit strong spin-orbit coupling and topological properties, the so-called Weyl semi-metals (see for instance Refs. \cite{armitage2017,Hasan_017,Yan_017,Burkov_016} and references therein). The first material candidate proposed as a realization of a Weyl semi-metal was the family of magnetic pyrochlore iridates $R_2Ir_2O_7$, where $R$ stands for a rare-earth element \cite{Wan_011}. Chirality is an intrinsically quantum mechanical property that, nonetheless, possesses macroscopic effects in transport phenomena on these materials.
Unlike in insulators, the non-trivial topology is linked to invariants associated to the Fermi surface rather than with completely filled bands in the bulk.
Despite some analogies with graphene due to the presence of Dirac points in the spectrum, the physics of topological Weyl semi-metals is richer and arises from the breaking of time reversal symmetry or inversion symmetry that leads to the splitting of a Dirac point into a pair of Weyl nodes of opposite chirality. On the one hand, they present spin-polarized surface states or Fermi arcs, that connect  the two bulk Weyl nodes of opposite chirality. On the other hand, despite their Dirac-like dispersion relation in the vicinity of the metallic points, they do not necessarily satisfy Lorentz covariance. Type II Weyl spinors
do actually violate this condition \cite{Hasan_017,armitage2017,Burkov_016,Yan_017}, by possessing a dispersion relation where the Dirac cone is strongly tilted with the consequent formation of electron and hole pockets \cite{Huang_016}. As important consequences of this effect, that are verifiable experimentally, are the chiral anomaly \cite{Morimoto_016} associated to a strong dependence on the direction of the applied electric field upon transport response, a modified anomalous Hall effect, and novel effects related to electron-electron interactions that may actually restore Lorentz invariance. One aspect of WSM that has not been explored in detail is the effect of mechanical strain on their transport properties. Although the effects of strain have been thoroughly studied in graphene (see for instance Refs. \cite{Amorim_016,Naumis-2017} and references therein), this is not the case for WSM. As in the case of graphene, different types of elastic strains give rise to different types of induced gauge fields in WSM \cite{Cortijo_015,Cortijo-2016,vozmediano2017-1}. These induced gauge fields can manifest themselves in the chiral anomaly effect\cite{Pikulin-2016}, in the energy spectrum \cite{Grushin-2016}, in quantum oscillations in the absence of a magnetic field \cite{Liu-2017} and in the collapse of the Landau levels \cite{Arjona-2017}. However, most of the analysis are done using tight-binding and numerical methods. In contrast, in this work we focus on the combined effects of torsional mechanical strain and an external magnetic field on the electronic transport properties in WSM, using an analytic procedure developed in Ref. \cite{Munoz-2017} by ourselves. Despite its relative simplicity, our model is able to demonstrate the node-filtering effect that can be achieved in WSMs under combined torsional strain and an external magnetic field. Moreover, we illustrate how this effect could be used in practice
for strain sensing in WSMs.

\section{Model} 
\label{sec:model}

In the massless case, the relativistic Dirac equation in three spatial dimensions (3D) can be written as two separate equations for two 2-component spinors of opposite helicity. The Weyl fermions are described by the $2\times2$ Hamiltonian:
\begin{equation}
\hat{H}_{0}^{\pm} =\pm c\vec{\sigma}\cdot\hat{\vec{p}}, 
\label{eq:relWeyl}
\end{equation}
where $\vec{\sigma}=(\sigma_1,\sigma_2,\sigma_3)$ are the three Pauli matrices, $c$ is the speed of light and $\hat{\vec{p}}$ is the momentum in 3D. As can be observed from eq. \eqref{eq:relWeyl} Weyl fermions propagate parallel (or antiparallel) to their spin, defining their chirality \cite{armitage2017}.

In the simplest case of a WSM with two nodes, the low energy effective theory is described by a block Hamiltonian involving both nodes (located at  $\mathbf{K}_{\pm}=\pm\vec{b}/2$, with $\vec{b}$ the vector between the two nodes in momentum space) \cite{vozmediano2017-1}:
\begin{equation}
\hat{H}_{b}=\bmb \hat{H}_{0}^{+} & 0\\
 0 & \hat{H}_{0}^{-} 
 \emb=v_F\bmb \vec{\sigma}\cdot\hat{\vec{p}} & 0\\
 0 & -\vec{\sigma}\cdot\hat{\vec{p}} 
 \emb
\end{equation}
where $v_F$ is the Fermi velocity. For instance, the physical values of $b$ and $v_F$ in TaAs \cite{Hasan_017} are $b\sim 0.08$ $\mathring{\text{A}}^{-1}$ and $v_F\sim 1.3\times 10^{-5}$ m/s. 

\subsection{Strain}
\label{sec:strain}

In a recent paper Arjona and Vozmediano \cite{vozmediano2017-1} show the appearance of pseudo-gauge fields associated to the presence of torsional strain. Closely following ref. [\onlinecite{vozmediano2017-1}] in this section we write the effective model for such a strain. In 3D the antisymmetric part of the deformation tensor is given by:
\begin{equation}
 \omega_{ij}=\frac{1}{2}\(\pd_i u_j-\pd_ju_i \)=\epsilon_{ijk}\Omega_k,
\end{equation}
where $\vec{\Omega}\equiv\frac{1}{2}\(\vec{\nabla}\times\vec{u}\right)$.
Let us focus now in the case of torsional strain applied to a cylindrical region of a WSM. Assuming that the two Weyl nodes are separated by a vector in the $z$-direction $\vec{b}=b\hat{z}$ and that the cylinder has a length $L$ in the $z$-direction, the displacement vector is given by \cite{vozmediano2017-1}: 
\begin{equation}
 \vec{u}=\theta\frac{z}{L}\(\vec{r}\times\hat{z}\)=\theta\frac{z}{L}\(y\hat{x}-x\hat{y}\).
\end{equation}
This vector gives rise to:
\begin{equation}
 \vec{\Omega}=\frac{1}{2}\(\vec{\nabla}\times\vec{u}\)=\frac{\theta}{2L}\(x\hat{x}+y\hat{y}-2z\hat{z}\),
\end{equation}
where the vector potential associated to the deformation is given by:
\begin{equation}
 \vec{A}=\vec{b}\times\vec{\Omega}=\frac{\theta b}{2L}\(-y\hat{x} +x\hat{y}\)
\end{equation}
and the associated induced pseudo-magnetic field is therefore:
\begin{equation}
 \vec{B}_s=\(\vec{\nabla}\times\vec{A}\)=\frac{\theta b}{L}\hat{z}.
 \label{eq:Bs}
\end{equation}
Let us notice that the pseudo-magnetic field will have different signs at each of the two nodes\cite{Cortijo_015,Cortijo-2016b}, while a real magnetic field has the same sign at both nodes.

Similar behavior can be observed in a different strain pattern as the one suggested by \cite{Arjona-2017}, where $u_x=u_0(2xy+C)$, $u_y=u_0\left[-x^2-Dy(y+C) \right]$ and $ u_z=0$. Nevertheless, our analysis will focus on the azimuthally symmetric torsional strain.  In addition to the torsional strain, we will also consider an external magnetic field that gives rise to the node-polarization effect on the electric current, similar to the valley-polarization effect observed in graphene \cite{Munoz-2017}.

\section{Scattering through a cylindrical region with magnetic field and mechanical strain}
\label{sec:scattering}

We start by considering the problem of three-dimensional elastic scattering of an incident free spinor with momentum $\mathbf{k} = (k_x,0,k_z)$ and energy $E_{k,\lambda}^{\xi} = \lambda \hbar v_F |\mathbf{k}|$, where $\lambda = \pm 1$ is the ``band'' index and $\xi = \pm 1$ labels each of the Weyl nodes located at $\mathbf{K}_{\xi}=\xi\vec{b}/2$.  Since $k_z$ is a good quantum number, it is possible to decouple the $z-$direction from the plane and therefore,  most of the analysis done in \cite{Munoz-2017} can be applied almost straightforwardly.

We consider a free fermion incident towards the cylindrical scattering center, described by the eigenvector of the equation
\begin{eqnarray}
\hat{H}_{0}^{\xi} \tilde{\Psi}_{in,\mathbf{k}}^{(\lambda,\xi)}(r,\phi,z) = \lambda \hbar v_F |\mathbf{k}| \tilde{\Psi}_{in,\mathbf{k}}^{(\lambda,\xi)}(r,\phi,z), 
\end{eqnarray}
which is given by the free spinor
\begin{equation}
\tilde{\Psi}_{in,\mathbf{k}}^{(\lambda,\xi)}(r,\phi,z) = \frac{1}{\sqrt{1 + \left(\frac{\lambda\xi|\mathbf{k}| - k_z}{k_x} \right)^2}}\left(\begin{array}{c}1 \\ \frac{\lambda\xi|\mathbf{k}| - k_z}{k_x} \end{array} \right) e^{i k_x r\cos\phi+ik_z z}.
\label{eq:incm}
\end{equation}
We now proceed with the standard partial wave analysis for scattering\cite{Munoz-2017,sakurai}. Using the symmetry around the $z-$axis, we choose as a quantum number the eigenvalue $\hbar m_j$ of the total angular momentum operator in the $z-$direction ($\hat{J}_3 = \hat{L}_3 + \hat{\sigma}_3/2$) to label the solutions:
\begin{equation}
\tilde{\Psi}_{m_j,k_z}^{(\lambda,\xi)}(r,\phi,z) = r^{-1/2}\left(\begin{array}{c} f_{m_j}(r) e^{i(m_j -1/2)\phi}\\-i\,g_{m_j}(r) e^{i(m_j + 1/2)\phi}
\end{array}\right)e^{ik_zz}. 
\label{eq:free}
\end{equation}

Closely following Ref. \cite{Munoz-2017} we found that:
\begin{align}
f_{m_j}(r) =& c_1 \sqrt{k_x r} J_{m_j -1/2}(k_x r) + c_2 \sqrt{k_x r} Y_{m_j - 1/2}(k_x r),\nonumber\\
g_{m_j}(r) =& c_3 \sqrt{k_x r} J_{m_j+1/2}(k_x r) + c_4 \sqrt{ k_x r} Y_{m_j+1/2}(k_x r),
\label{eq:bessel}
\end{align}
with
\begin{eqnarray}
c_3 = -\frac{k_x}{k_z+\lambda\xi |\mathbf{k}|}c_1,\,\,\,\,
c_4 = -\frac{k_x}{k_z+\lambda\xi |\mathbf{k}|}c_2.
\end{eqnarray}

 In addition, as in standard elastic scattering theory \cite{sakurai}, the phase shift captures the effect of a scattering region over the transmitted particle waves.  

In order to calculate the phase shift $\delta_m$ associated to each angular momentum channel $m \equiv m_j - 1/2$, it is necessary to match each spinor component of the free solution Eq. \eqref{eq:free}, and its first derivative, to the solution inside the region submitted to the effective magnetic field $B_\xi=B_0+\xi B_S$ (where $B_0$ is the external magnetic field and $B_s=\theta b/L$ is the pseudo-magnetic field induced by the torsional strain, Eq. \eqref{eq:Bs}) at the boundary of the cylinder $r=a$. The spectrum inside the cylindrical region corresponds to relativistic Landau levels, as seen in Ref.~\cite{Munoz-2017}, with an effective magnetic field that is node-dependent,
\begin{eqnarray}
E_{\lambda}^{\xi}(n) = \lambda \hbar v_F \sqrt{2 n |B_{\xi}|/\tilde{\phi}_0  + k_z^2}, 
\label{eq_spectrum}
\end{eqnarray}
with $\tilde{\phi}_0 = \hbar v_F/e$ the flux quantum defined in terms of the Fermi velocity in the WSM.
As it is explained in detail in Ref. \cite{Munoz-2017}, one can find an exact analytical expression for the phase shift $\delta_m$,
\begin{widetext}
\begin{equation}
\tan\delta_m = -\frac{c_2}{c_1}= \frac{J_{m+1}(k_x a) + \displaystyle\frac{J_{m}(k_xa)}{k_x a}\left\{|m| - m - \frac{|B_{\xi}|a^2}{2\tilde{\phi}_0}- \frac{L_{n_{\rho}-1}^{|m|+1}(|B_{\xi}|a^2/2\tilde{\phi}_0)}{L_{n_{\rho}}^{|m|}(|B_{\xi}|a^2/2\tilde{\phi}_0)}
\right\}}
{Y_{m+1}(k_x a) + \displaystyle\frac{Y_{m}(k_x a)}{k_x a}\left\{
|m| - m - \frac{|B_{\xi}|a^2}{2\tilde{\phi}_0}- \frac{L_{n_{\rho}-1}^{|m|+1}(|B_{\xi}|a^2/2\tilde{\phi}_0)}{L_{n_{\rho}}^{|m|}(|B_{\xi}|a^2/2\tilde{\phi}_0)}
\right\}}.
\label{eq:phase}
\end{equation}
\end{widetext}
where we have defined $m \equiv m_j - 1/2$.
 
\subsection{Scattering cross section}
Far from the cylindrical region where the effective magnetic field is present ($r\gg a$), the state is a linear combination of the incident and scattered spinor:
\begin{equation}
\tilde{\Psi}_{out}(r,\phi,z) \sim\frac{1}{\sqrt{1 + \left(\frac{\lambda\xi|\mathbf{k}| - k_z}{k_x} \right)^2}}\left(\begin{array}{c}1 \\ \frac{\lambda\xi|\mathbf{k}| - k_z}{k_x} \end{array} \right) e^{i k_x r\cos\phi+ik_ z} + \left( \begin{array}{c}f_{1}(\phi)\\f_{2}(\phi)\end{array} \right)\frac{e^{ik_x r + i k_z  z}}{\sqrt{r}},
\label{eq:out}
\end{equation}
with amplitudes $f_1(\phi)$ and $f_2(\phi)$ for each component of the scattered spinor.
 In the same region, we have that this expression must be equal to the asymptotic form of the solution, represented in terms of
phase shifts as explained in Ref. \cite{Munoz-2017}. We find the following scattering amplitudes:

\begin{equation}
\left( \begin{array}{c} f_1(\phi)\\f_2(\phi)\end{array}\right) = \frac{e^{-i\pi/4}}{\sqrt{2 \pi k_x}\sqrt{1 + \left(\frac{\lambda\xi|\vec{k}|-k_z}{k_x} \right)^2}}\sum_m\left(\begin{array}{c} e^{im\phi} \\
\frac{k_x}{\lambda\xi |\vec{k}| + k_z} e^{i(m+1)\phi}\end{array} \right)\left( e^{2 i \delta_m} -1 \right).
\label{eq:amp}
\end{equation}

For a very long cylinder, $L \gg 1/k_F$, the differential scattering cross-section per unit length is given by the modulus of the vector above,
\begin{equation}
\frac{d\tilde{\sigma}}{d\phi} = |f_1(\phi)|^2 + |f_2(\phi)|^2 = \frac{2}{\pi k_x}\frac{|\vec{k}|+\lambda\xi k_z}{|\vec{k}| - \lambda\xi k_z}\sum_{m,m'} e^{i(\delta_m - \delta_{m'})}\sin\delta_m \sin\delta_{m'} e^{i(m - m')\phi}
\label{eq:cross_section}
\end{equation}
and the total scattering cross section is then given by integrating over the scattering angle $\phi$ ($0\le \phi \le 2\pi$) and over the length of
the cylinder
\begin{align}
\sigma =& \int_{0}^{L} dz \int_0^{2\pi} d\phi\, \frac{d\tilde{\sigma}}{d\phi} = L \int_0^{2\pi} \left(  |f_1(\phi)|^2 + |f_2(\phi)|^2  \right)d\phi\nonumber\\
=& \frac{4 L}{k_x}\frac{|\vec{k}|+\lambda\xi k_z}{|\vec{k}| - \lambda\xi k_z}\sum_{m=-\infty}^{\infty}\sin^2\delta_{m}.
\label{eq_total_scatt}
\end{align}

Notice that the assumption of a very long cylinder, $L \gg 1/k_F$, is well founded. For instance, in TaAs where $b\sim 0.08$ $\mathring{\text{A}}^{-1}$ and $v_F\sim 1.3\times 10^{5}$ m/s, $1/k_F\sim 9$ $\mathring{\text{A}}$, so even a slab of a few microns is already in the range of validity of our expressions. 
Moreover, for Cd$_3$As$_2$, $b\sim 0.2$ $\mathring{\text{A}}^{-1}$ and $v_F\sim 1.5\times 10^{6}$ m/s, $1/k_F\sim 0.8$ $\mathring{\text{A}}$ \cite{neupane-2014}
and hence the applicability of our expressions is even more striking in this second example.

\begin{figure}[hbt]
\centering
\includegraphics[width=0.6\columnwidth]{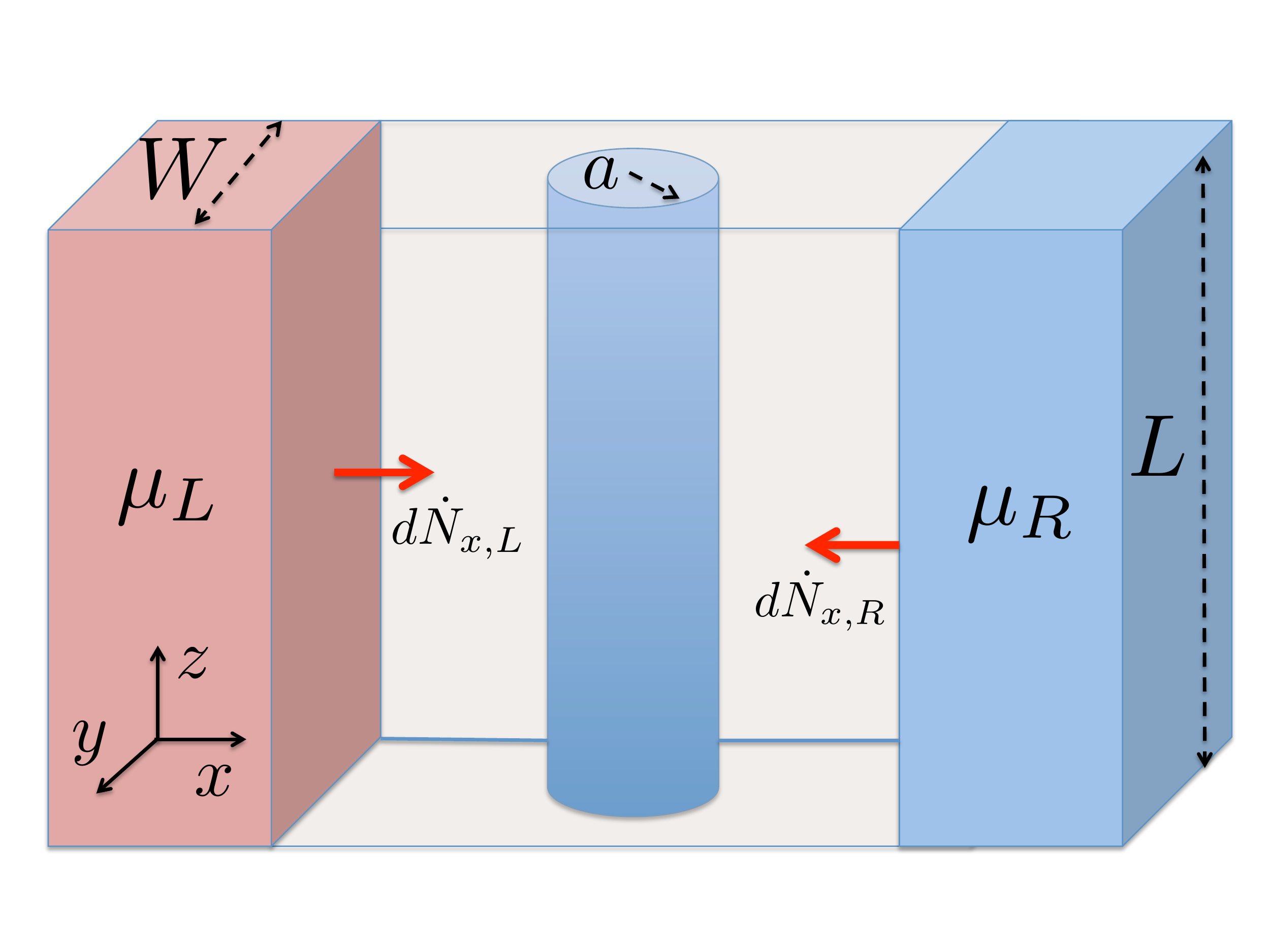}
\caption{(Color online) Pictorial description of the system under consideration: A WSM slab of dimensions $L\times W$, with a cylindrical region of radius $a$ submitted to a combination of torsional strain and an external magnetic field $\vec{B}=B_0\hat{z}$.}
\label{fig1}
\end{figure}

\section{Transmission and Landauer ballistic current}

Let us consider a slab of a WSM of high $L$ and width $W$ ($y$-direction), which is connected to two semi-infinite WSM contacts held at chemical potentials $\mu_L$ and $\mu_R$, respectively. In addition, we consider that inside a cylindrical region of radius $a$ a uniform magnetic field in the $z-$direction (along the cylinder axis) is applied ($B_0\hat{z}$). Moreover, we assume that in the same cylindrical region a torsional mechanical strain is applied, giving rise to an induced pseudo-magnetic field in the $z-$direction as it was mentioned in sec. \ref{sec:strain}. Using the Landauer ballistic approach, the net current along the slab ($x$-direction) is given by the net counterflow of the particle currents emitted from the left and right semi-infinite WSM contacts, respectively. Each contact is assumed to be in thermal equilibrium, with the Fermi-Dirac distributions $f(E-\mu_L,T)\equiv f_L(E)$ and  $f_R(E) \equiv f(E-\mu_R,T)$, respectively.  A pictorial description of the system is shown in Fig.~\ref{fig1}.

The particle flux emitted by the left (L) and right (R) contacts, respectively, is defined as
\begin{equation}
dJ_{x,L/R} = v_x D_{L/R}(E) f_{L/R}(E) dE,
\end{equation}
where $D_{L/R}(E)$ is the density of states at each contact.

The effect of the external magnetic field and the pseudo-magnetic field over charge transport can be expressed as an effective  cross-section $W L T_{\xi}(E,\phi)$, with $T_{\xi}(E,\phi)$ the transmission coefficient in the direction specified by the polar angle $\phi$, for an incident spinor arising from the node $\mathbf{K}_{\xi}$. We thus define the effective cross-section in the $\phi$-direction by:
\begin{align}
W L\, T_{\xi}(E_k,\phi) &= \sum_{n,\lambda}\frac{L}{\tilde{\sigma}(\mathbf{k})} \frac{d\tilde{\sigma}(\mathbf{k})}{d\phi}  \delta\left(\lambda |\vec{k}| - \frac{E_{\lambda}^{\xi}(n)}{\hbar v_F}\right),
\label{eq_transm}
\end{align}
with $E_{\lambda}^{\xi}(n)$ the energy spectrum inside the cylindrical region, as defined in Eq.(\ref{eq_spectrum}). We will take from now on that the incident spinor has a wave vector $\vec{k}=(k_x,0,0)$, i.e. a plane wave propagating in the $x-$direction. The particle flow (per unit time) along the $x$-direction emitted by the left L (right R) contact and arising from the $\mathbf{K}_{\xi}$ node can be written as $d\dot{N}_{x,L(R)}^{\xi} = W L\,T_{\xi}(E,\phi)d\phi\, dJ_{x,L(R)}$, where $v_x = v_F\cos\phi$. The net electric current flowing across the region will be $I = I_{+} + I_{-}$, with the node component given by 
\begin{eqnarray}
I_{\xi} = e\int\left( d\dot{N}_{x,L}^{\xi} - d\dot{N}_{x,R}^{\xi} \right)= e v_F W L  \int_{-\infty}^{\infty} dE \left[ D_L(E)f_L(E) - D_R(E)f_R(E)\right] \bar{T}_{\xi}(E).
\label{eq_current0}
\end{eqnarray}

Here,  we have defined the net transmission coefficient for Dirac spinors at node $\mathbf{K}_{\xi}$ as the angular average $\bar{T}_{\xi}(E) = \int_{-\pi/2}^{\pi/2}d\phi\,\cos\phi\,T_{\xi}(E,\phi)$.

Assuming that both contacts are identical semi-infinite regions, the density of states are equal, and given by
\begin{eqnarray}
D_L(E) &=& D_R(E) = \frac{2}{\pi(\hbar v_F)^2L}\left|E\right| \theta(|E|),
\end{eqnarray}
where the factor of $4$ arises from the spin and node degeneracy at each of the WSM semi-infinite contacts.
With this consideration, the expression for the node component of the current $I_{\xi}$ becomes
\begin{align}
I_{\xi} 
&=\frac{8 e v_F }{\pi^2}\sum_{\lambda,n,m,p}\frac{(-1)^{p+1}}{\tilde{\sigma}(E_{\lambda}^{\xi}(n))(4p^2-1)}
e^{i(\delta_m - \delta_{m-2p})}\sin\delta_m\sin\delta_{m-2p}\left[ f_L(E_{\lambda}^{\xi}(n)) - f_R(E_{\lambda}^{\xi}(n))\right],
\label{eq_current}
\end{align}
with the total current given by $I = I_{+} + I_{-}$.

\section{Results and Discussion}

\begin{figure}[hbt]
\centering
\subfigure[]{\includegraphics[width=0.322\textwidth]{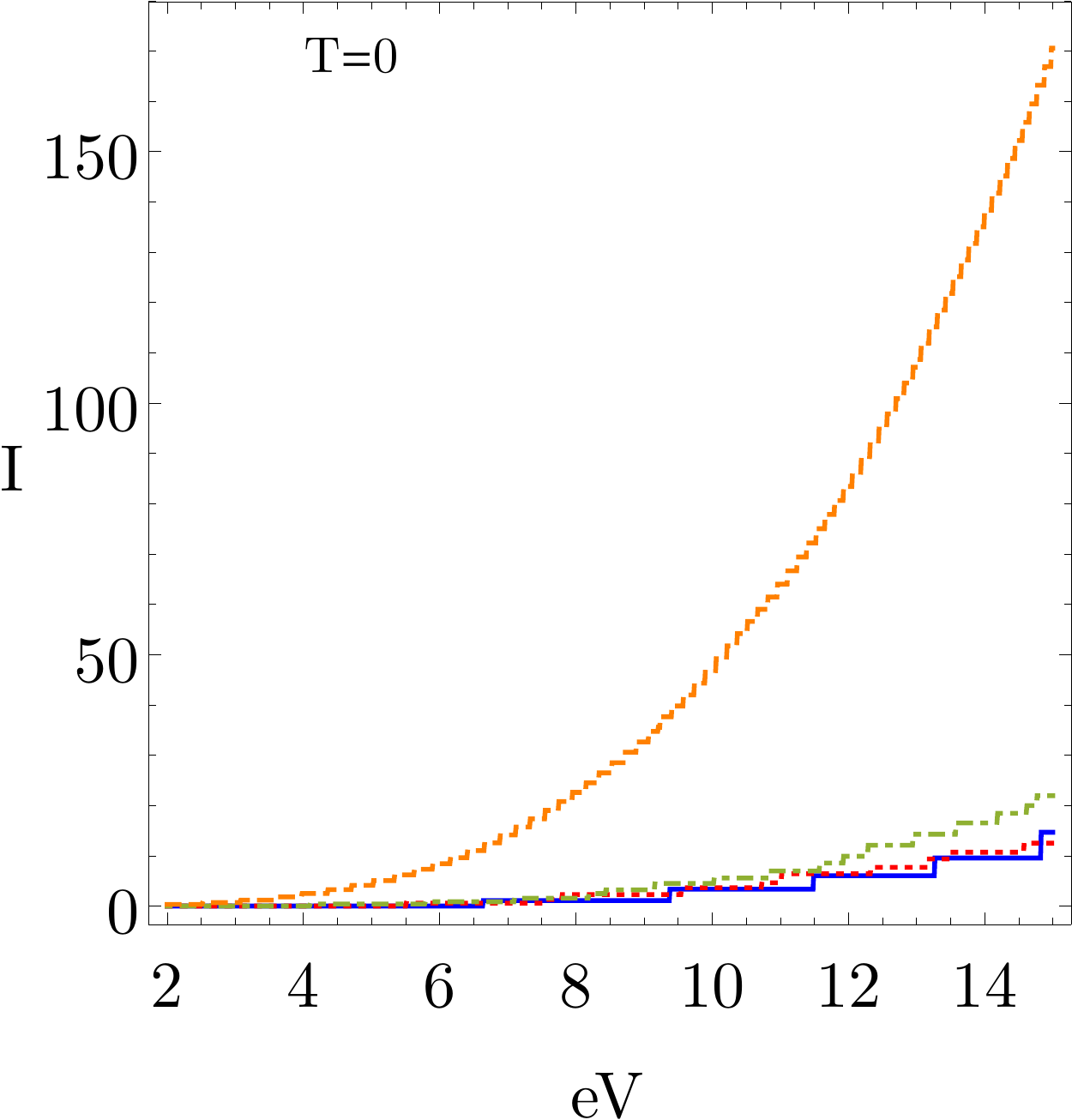}\label{fig2a} }
\hskip .1cm 
\subfigure[]{\includegraphics[width=0.322\textwidth]{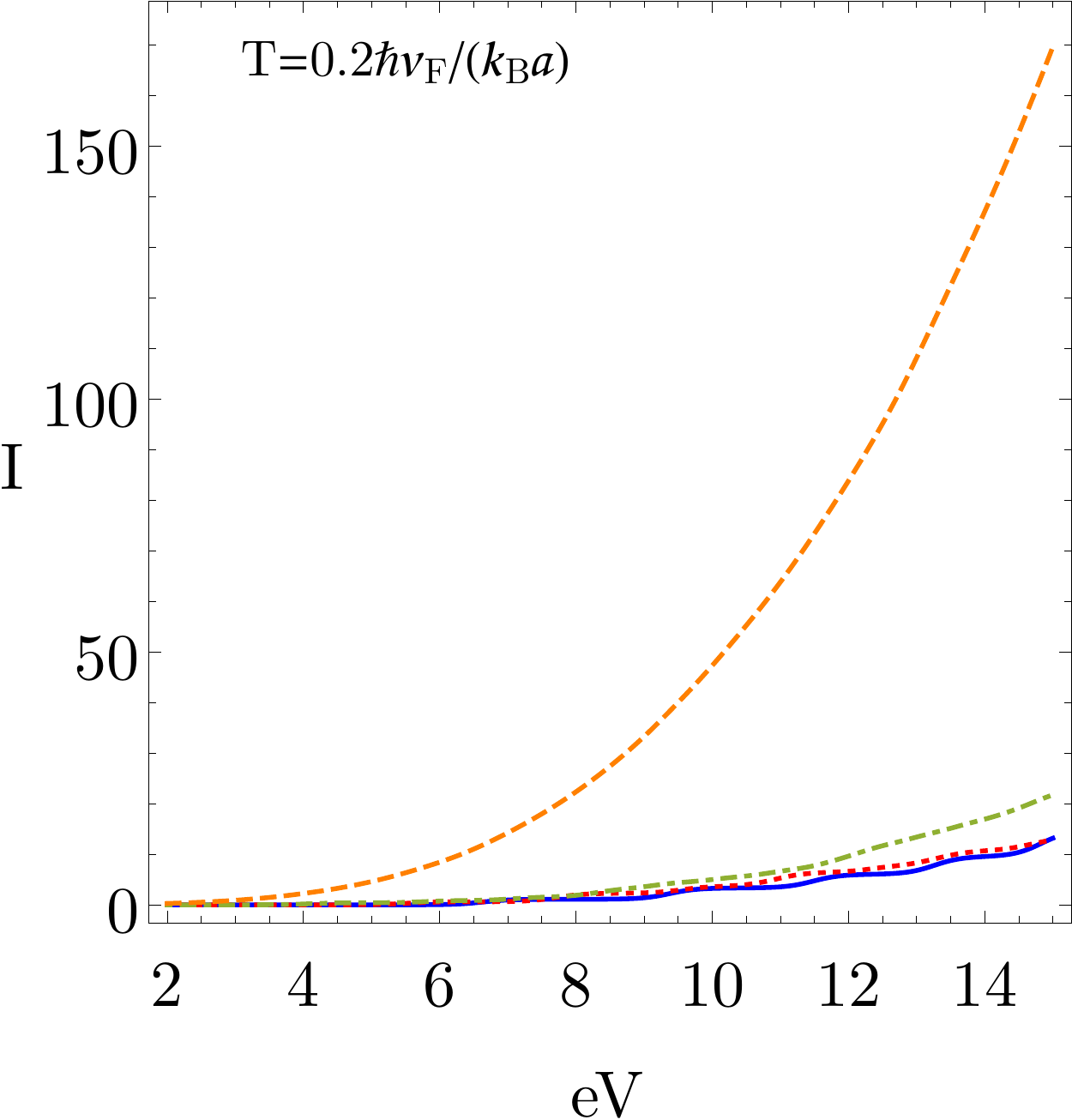}\label{fig2b}}
\hskip .1cm 
\subfigure[]{\includegraphics[width=0.322\textwidth]{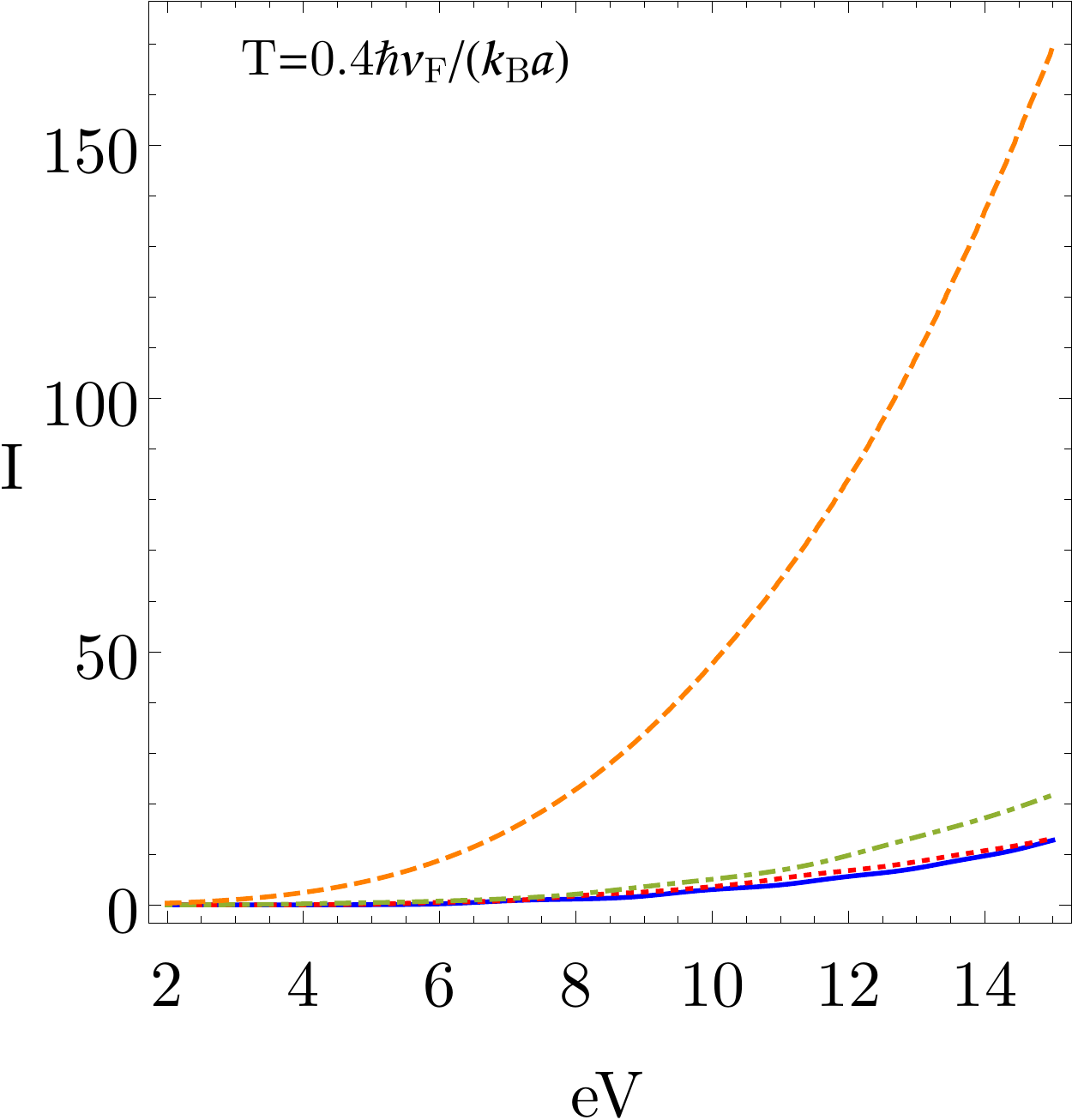}\label{fig2c}}
\caption{(Color online) Current (in units of $e v_F/a$) calculated from the analytical Eq.(\ref{eq_current}), as a function of applied bias $V$ (in units of $\hbar v_F/a$), for fixed $B_0a^2=22\tilde{\phi}_0$ and different values of the torsion angle $\theta$. The solid (blue) line corresponds to $\theta=0$ ($B_Sa^2=0$), the dotted (red) line corresponds to $\theta=5^\circ$ ($B_Sa^2=6.8\tilde{\phi}_0$), the dotdashed (green) line corresponds to $\theta=10^\circ$ ($B_Sa^2=13.6\tilde{\phi}_0$) and the dashed (orange) line corresponds to $\theta=15^\circ$($B_Sa^2=20.4\tilde{\phi}_0$), with $\tilde{\phi}_{0}\equiv \hbar v_F /e$. The subfigures (a), (b) and (c) correspond to the different values of the temperature, $T=0$, $T = 0.2 \,\hbar v_F/ (k_B a)$ and $T = 0.4 \,\hbar v_F/ (k_B a)$ respectively.}
\label{fig2}
\end{figure}

\begin{figure}[hbt]
\centering
\subfigure[]{\includegraphics[width=0.322\textwidth]{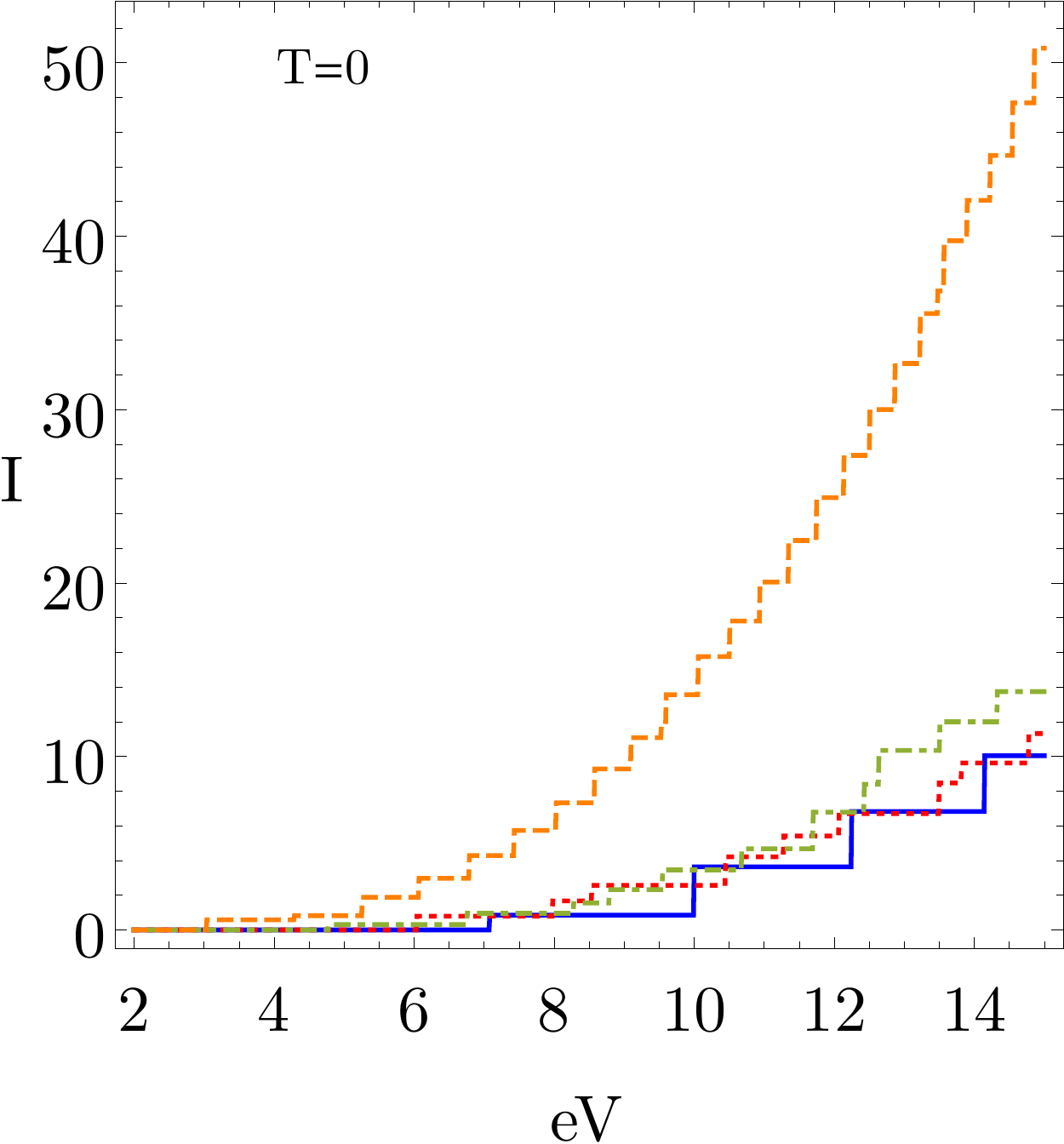}\label{fig21a} }
\hskip .1cm 
\subfigure[]{\includegraphics[width=0.322\textwidth]{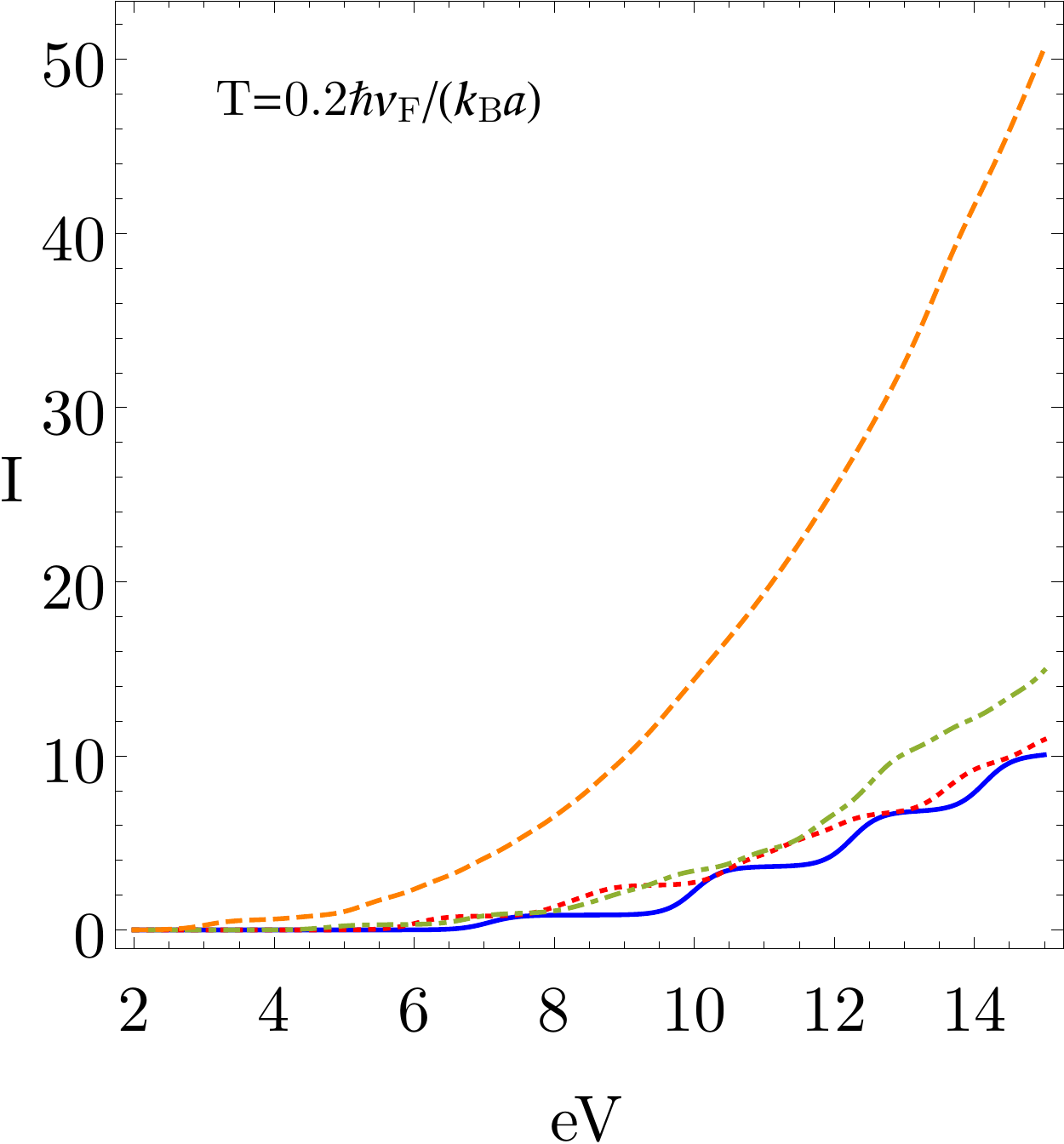}\label{fig21b}}
\hskip .1cm 
\subfigure[]{\includegraphics[width=0.322\textwidth]{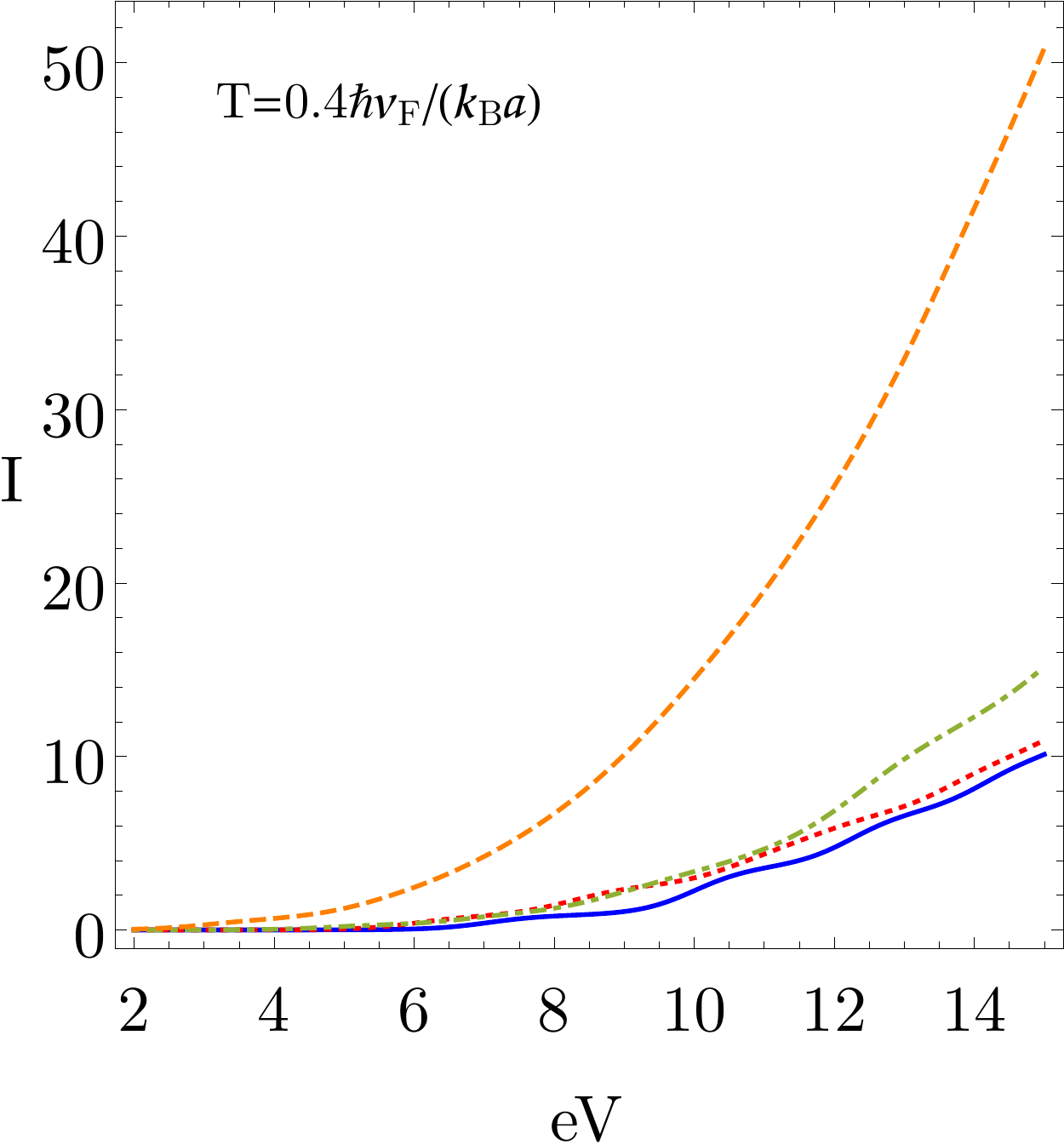}\label{fig21c}}
\caption{(Color online) Current (in units of $e v_F/a$) calculated from the analytical Eq.(\ref{eq_current}), as a function of applied bias $V$ (in units of $\hbar v_F/a$), for fixed $B_0a^2=25\tilde{\phi}_0$ and different values of the torsion angle $\theta$. The solid (blue) line corresponds to $\theta=0$ ($B_Sa^2=0$), the dotted (red) line corresponds to $\theta=5^\circ$ ($B_Sa^2=6.8\tilde{\phi}_0$), the dotdashed (green) line corresponds to $\theta=10^\circ$ ($B_Sa^2=13.6\tilde{\phi}_0$) and the dashed (orange) line corresponds to $\theta=15^\circ$($B_Sa^2=20.4\tilde{\phi}_0$), with $\tilde{\phi}_{0}\equiv \hbar v_F /e$. The subfigures (a), (b) and (c) correspond to the different values of the temperature, $T=0$, $T = 0.2 \,\hbar v_F/ (k_B a)$ and $T = 0.4 \,\hbar v_F/ (k_B a)$ respectively.}
\label{fig2_1}
\end{figure}

\begin{figure}[hbt]
\centering
\subfigure[]{\includegraphics[width=0.322\textwidth]{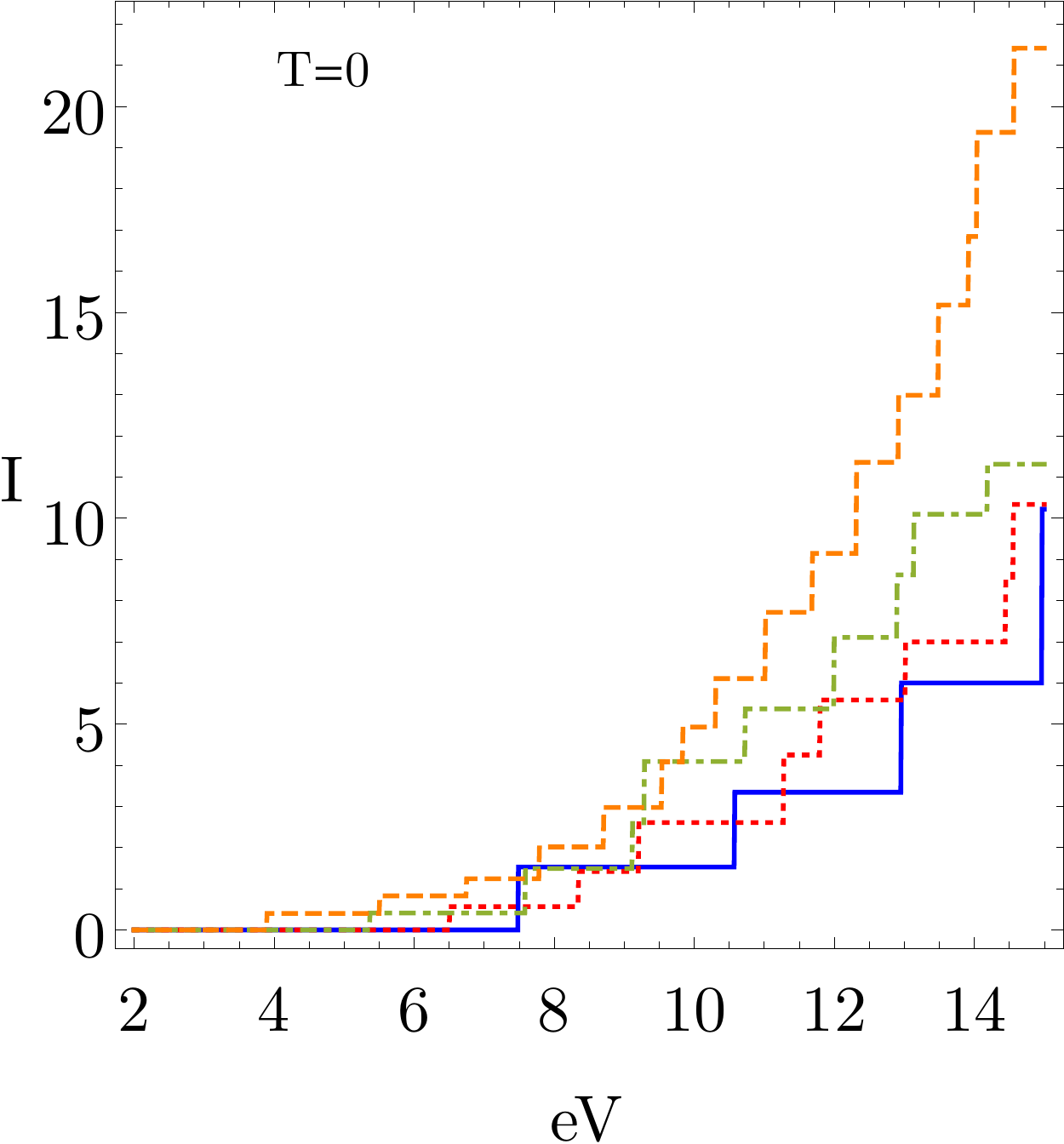}\label{fig22a} }
\hskip .1cm 
\subfigure[]{\includegraphics[width=0.322\textwidth]{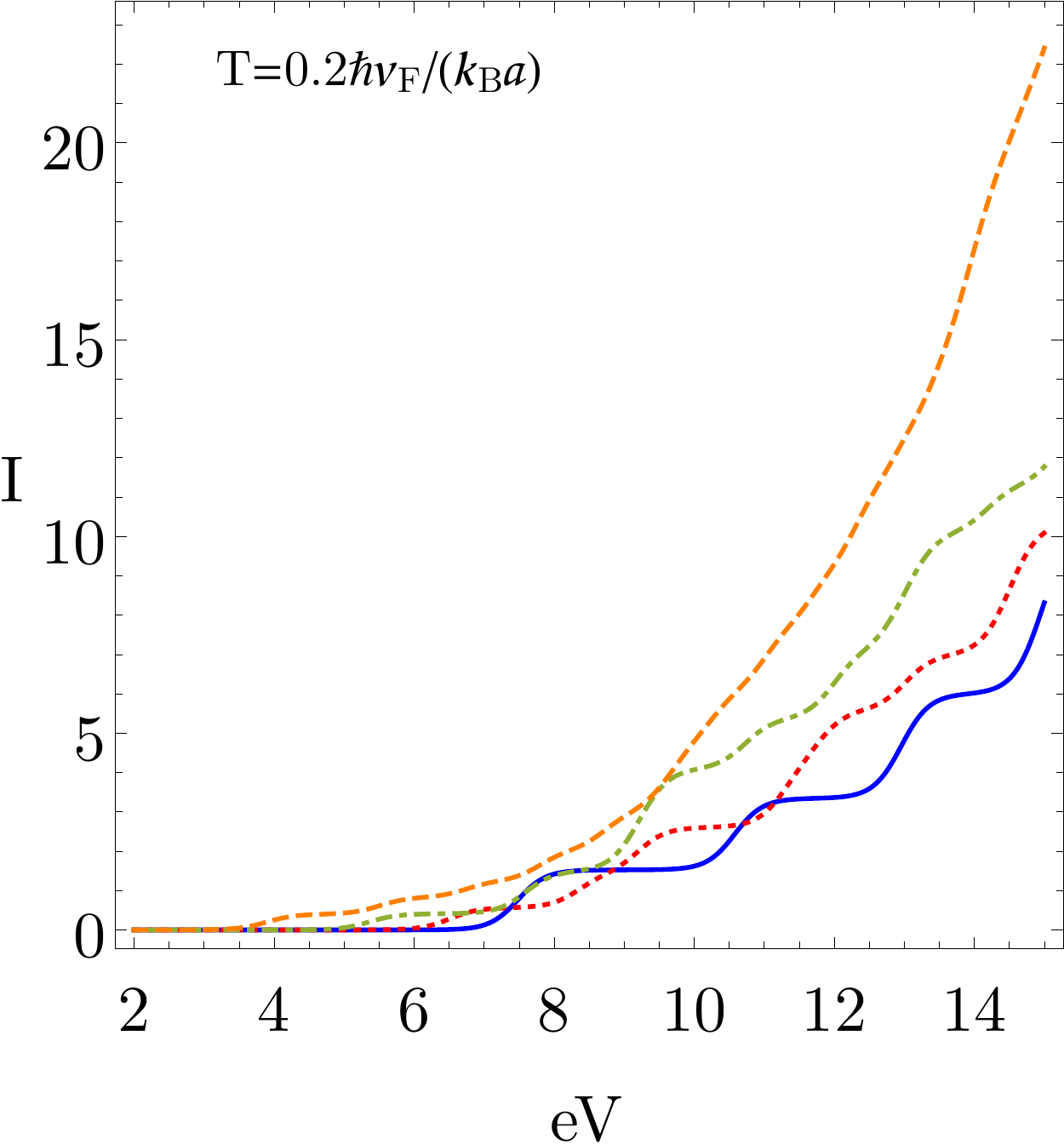}\label{fig22b}}
\hskip .1cm 
\subfigure[]{\includegraphics[width=0.322\textwidth]{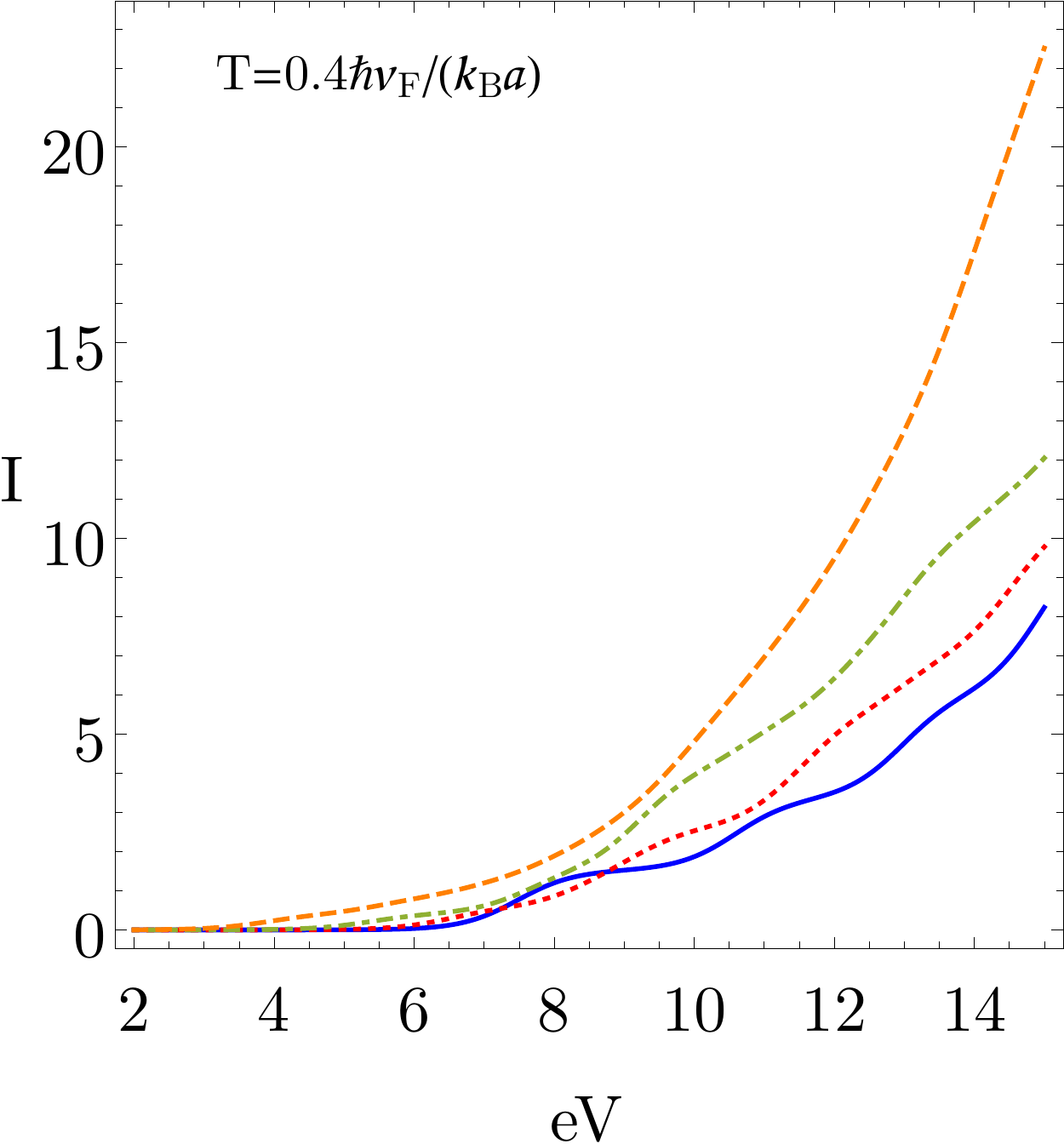}\label{fig22c}}
\caption{(Color online) Current (in units of $e v_F/a$) calculated from the analytical Eq.(\ref{eq_current}), as a function of applied bias $V$ (in units of $\hbar v_F/a$), for fixed $B_0a^2=28\tilde{\phi}_0$ and different values of the torsion angle $\theta$. The solid (blue) line corresponds to $\theta=0$ ($B_Sa^2=0$), the dotted (red) line corresponds to $\theta=5^\circ$ ($B_Sa^2=6.8\tilde{\phi}_0$), the dotdashed (green) line corresponds to $\theta=10^\circ$ ($B_Sa^2=13.6\tilde{\phi}_0$) and the dashed (orange) line corresponds to $\theta=15^\circ$($B_Sa^2=20.4\tilde{\phi}_0$), with $\tilde{\phi}_{0}\equiv \hbar v_F /e$. The subfigures (a), (b) and (c) correspond to the different values of the temperature, $T=0$, $T = 0.2 \,\hbar v_F/ (k_B a)$ and $T = 0.4 \,\hbar v_F/ (k_B a)$ respectively.}
\label{fig2_2}
\end{figure}

\begin{figure}[hbt]
\centering
\includegraphics[width=0.45\textwidth]{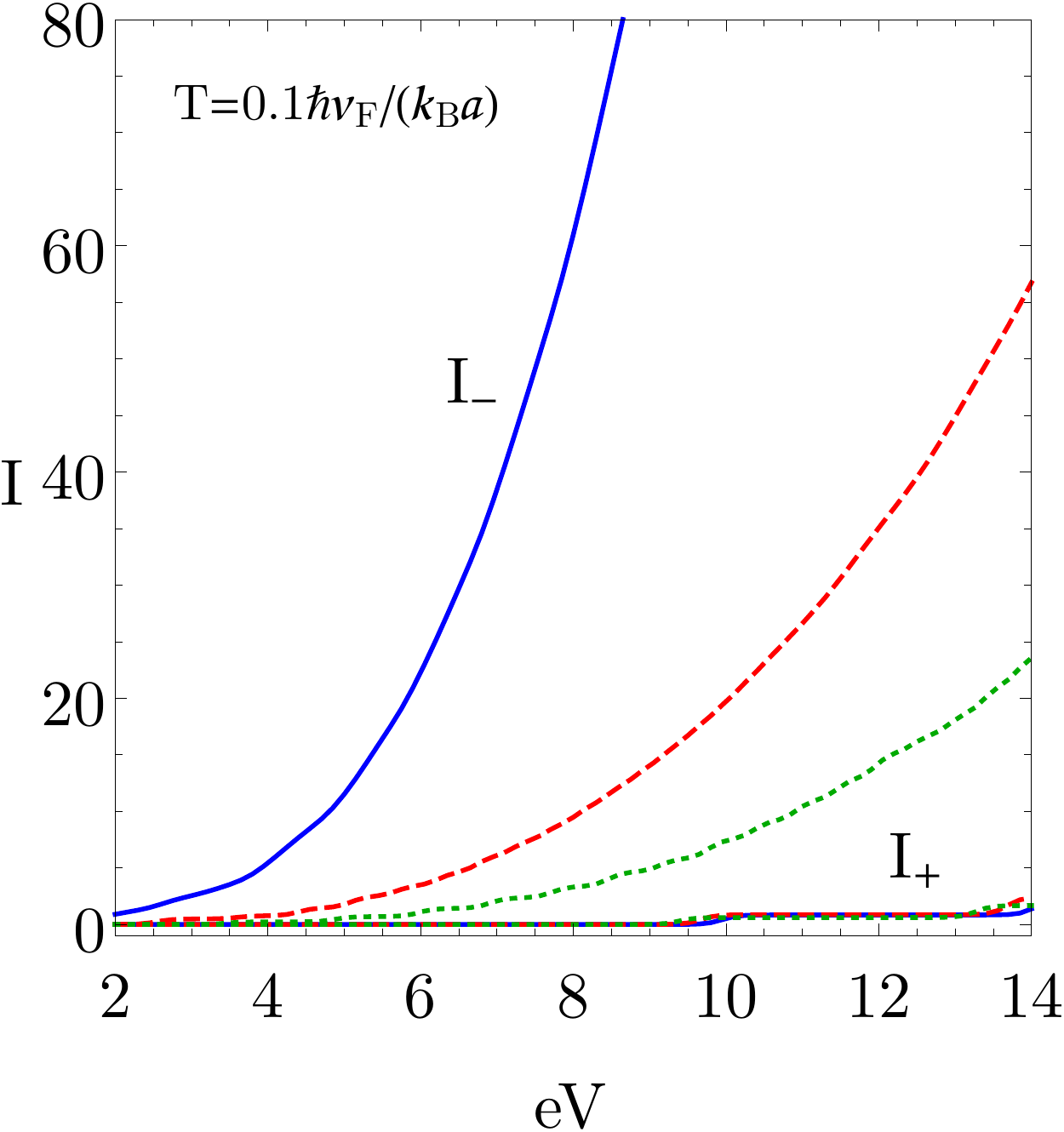}
\caption{(Color online) Node-polarized components of the current (in units of $e v_F/a$), calculated from the analytical Eq.(\ref{eq_current}), as a function of the applied bias $e V$ (in units of $\hbar v_F/a$), at finite temperature $T=0.1\,\hbar v_F/ (k_B a)$, fixed $B_0a^2=25\tilde{\phi}_0$ and different  values of the torsion angle $\theta$.
The solid (blue) lines correspond to $\theta=18^\circ$ ($B_Sa^2=24.5\tilde{\phi}_0$), the dashed (red) lines correspond to  $\theta=16^\circ$ ($B_Sa^2=21.8\tilde{\phi}_0$) and  the dotted (green) lines correspond to $\theta=14^\circ$ ($B_Sa^2=19\tilde{\phi}_0$), with $\tilde{\phi}_{0}\equiv \hbar v_F/e$. 
}
\label{fig3}
\end{figure}

\begin{figure}[hbt]
\centering
\subfigure[]{\includegraphics[width=0.4277\textwidth]{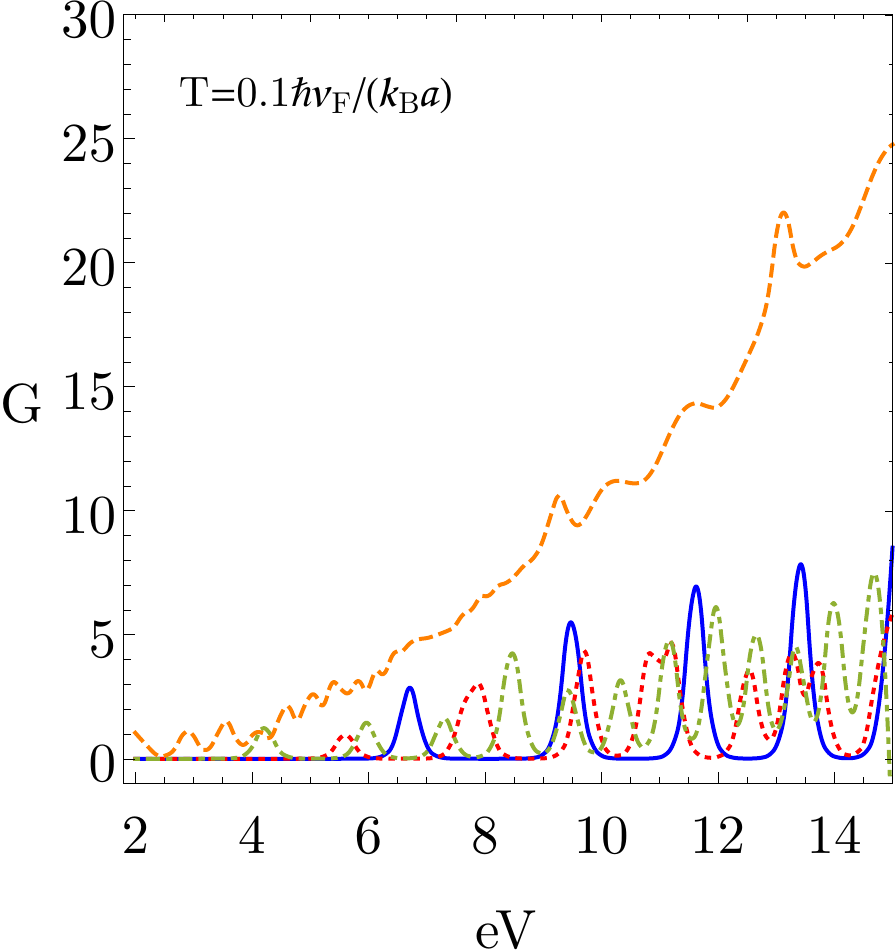}\label{fig4a}}
\subfigure[]{\includegraphics[width=0.4277\textwidth]{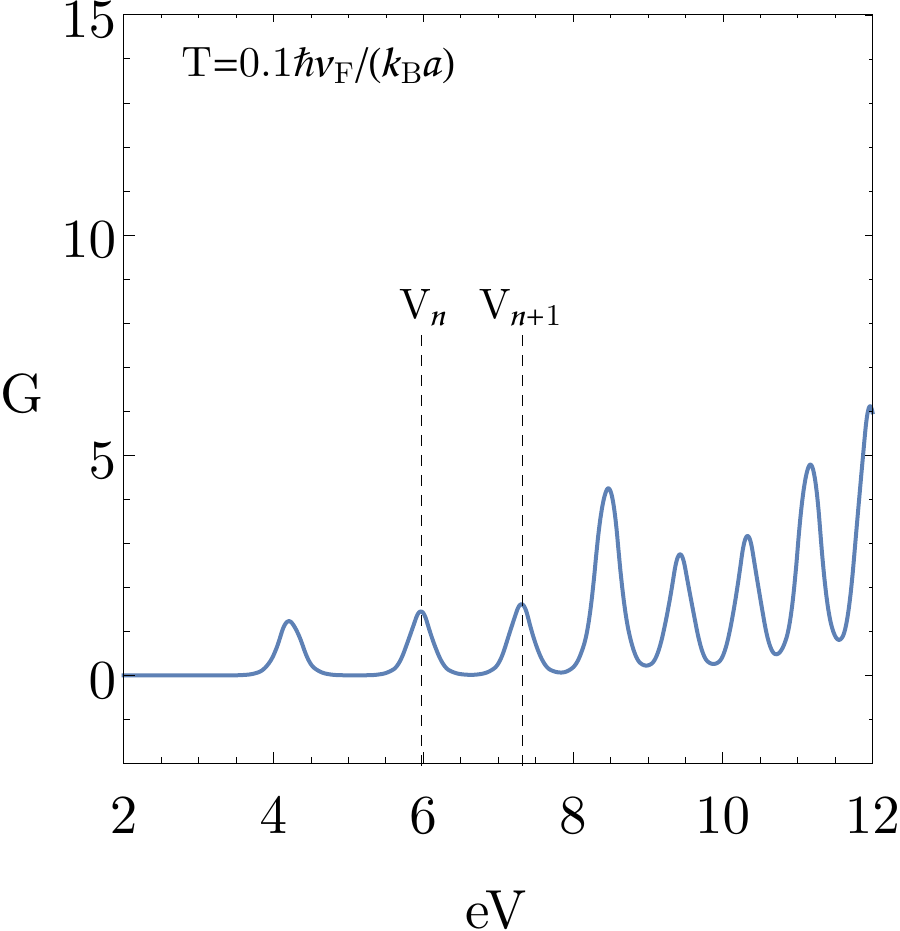}\label{fig4b}}
\caption{(Color online)(a) Conductance (in units of $e^2/\hbar$) as a function of bias $e V$ (in units of $\hbar v_F/a$), calculated as the voltage-derivative of the analytical Eq.(\ref{eq_current}), for fixed $B_0a^2=22.5\tilde{\phi}_0$ and different values of the torsion angle $\theta$. The solid (blue) line corresponds to $\theta=0$, the dotted (red) line corresponds to $\theta=5^\circ$, the dotdashed (green) line corresponds to $\theta=10^\circ$ and the dashed (orange) line corresponds to $\theta=15^\circ$, with $\tilde{\phi}_{0}\equiv \hbar v_F /e$. (b) Conductance (in units of $e^2/\hbar$) as a function of bias $e V$ (in units of $\hbar v_F/a$) for fixed $B_0a^2=22.5\tilde{\phi}_0$ and $\theta=10^\circ$ at a temperature $T = 0.1 \,\hbar v_F/ (k_B a)$}
\label{fig5}
\end{figure}

In analogy to the case of graphene discussed in \cite{Munoz-2017}, we calculate in this section the total current $I = I_{+} + I_{-}$ (in units of $ev_F/a$) Eq.~\eqref{eq_current} for the two nodes
components $I_{\xi}$, at zero and finite temperatures (Figs. \ref{fig2a} - \ref{fig2c}, \ref{fig21a} - \ref{fig21c} and \ref{fig22a} - \ref{fig22c}), for the particular choice of contact chemical potentials $\mu_L = eV$ and $\mu_R = 0$. 

Pikulin and collaborators \cite{Pikulin-2016} estimated the value of the induced magnetic field $B_S\approx1.8\times 10^{-3}T$ per angular degree of torsion for a cylinder of length $L\sim 100$ nm and radius $a\sim 50$ nm. In addition, we define the scaled flux quantum $\displaystyle\tilde{\phi}_{0}\equiv \frac{\hbar v_F}{e}=\frac{1}{2\pi}\frac{v_F}{c}\frac{hc}{e}=\frac{1}{2\pi}\frac{1.5}{300}\cdot4.14\times10^5$ T$\mathring{\text{A}}^2\approx330$ T$\mathring{\text{A}}^2$. Using these values we can write:
\begin{equation}
 B_Sa^2=1.36\theta \tilde{\phi}_{0},
 \label{eq:torsionangle}
\end{equation}
where $\theta$ is the torsion (twist) angle in degrees. In the following, we plot the current and conductance for different values of the torsion angle $\theta$ and different values of the external magnetic field.

In Fig.\ref{fig2_1}(a) we show the total current for $T = 0$, as a function of the applied bias voltage $V$, for a fixed value of the external magnetic field $B_0a^2 =25\tilde{\phi}_{0}$. The different curves correspond to different values of the torsion angle $\theta$, which gives different values for the  induced pseudo-magnetic field $B_S$. We can see that different values of the torsion angle produce different current-voltage curves. An important feature of these curves is the appearance of plateaus, which are more manifest at lower values of strain. This effect can be understood from the elastic scattering (see Eq.\eqref{eq_transm}) restriction, which gives the condition that for the incident particle to be transmitted across the scattering region its energy must be resonant with one of the eigenstates in the cylindrical region submitted to the effective magnetic field. 
The quasi-continuum distribution of energy values in the contacts allows for this condition be always
fulfilled, for an interval within the window imposed by the external bias voltage. Although the plateaus can still be seen at low finite temperature (see Fig. \ref{fig21b}), they are smoother than at zero temperature and they eventually disappear at high enough temperatures, as can be appreciated in Fig \ref{fig21c}.

In Fig.\ref{fig2} at $B_{0}a^2 = 22\tilde{\phi}_0$, Fig.\ref{fig2_1} at $B_{0}a^2 = 25\tilde{\phi}_0$ and Fig.\ref{fig2_2} at $B_{0}a^2 = 28\tilde{\phi}_0$, we compare the effect of the external magnetic field $B_0$ on the current-voltage curves. At zero and finite temperatures, one can observe that for the same values of torsion angle, i.e. $\theta = (5^\circ,10^\circ,15^\circ)$, the total current decreases as the external magnetic field is increased from $B_{0}a^2 = 22\tilde{\phi}_0$ (in Fig.\ref{fig2}) towards $B_{0}a^2 = 28\tilde{\phi}_0$ (in Fig.\ref{fig2_2}). This effect can be understood looking at the density of the pseudo-Landau level spectrum at the nodes. For a fixed value of $B_S$ the effective pseudomagnetic field $|B_{\xi}| = |B_{0} + \xi B_S|$ increases as $B_0$ increases (we are taking $B_0>B_S$ in Figures \ref{fig2}-\ref{fig2_2}). This produces a lower spectral density at the nodes $E_n^{(\xi)} \sim \sqrt{|B_{\xi}|n}$, giving less channels to the incident particles to be transmitted and decreasing the total current $I$ across the contacts.

Another feature of the current-voltage curves that is important to mention is the relative enhancement of the node-polarized contribution arising from the $\mathbf{K}_{-}$ node, as can be clearly observed in Fig.\ref{fig3}, where the two nodes components $I_{+}$ and $I_{-}$ of the total current are represented at finite temperature. This effect is even stronger when $B_S$ is closer to $B_0$ and can be understood by
noticing that the effective pseudomagnetic field $|B_{-}| = |B_{0} - B_S|$ decreases as $B_S$ increases, thus leading to a higher spectral density associated to the $\mathbf{K}_{-}$-node where $E_n^{(-)} \sim \sqrt{|B_{-}|n}$. This gives more channels to the incident particles to be transmitted per finite interval of bias voltage imposed at the contacts, thus increasing the current $I_{-}$ associated to the $\mathbf{K}_{-}$-node. This effect can also be appreciated at finite temperatures, as
seen in Fig.~\ref{fig3}, and it could be used to design a node-sensitive filter.

The differential conductance $G(V,T) = dI/dV$ (in units of $e^2/\hbar$) at 
finite temperature $T = 0.1 \,\hbar v_F/ (k_B a)$ is displayed in Fig.~\ref{fig4a} for different values of  the torsion angle $\theta$ and fixed $B_0=22.5\tilde{\phi}_0/a^2$. A characteristic trend of oscillations is observed, which are consistent with the staircase behavior of the current observed in Figs.~\ref{fig2}--\ref{fig2_2}

\subsection{Torsional Strain Sensing}

As it was explained in Ref. \cite{Munoz-2017} for the case of graphene, the dependence of the current-voltage characteristics on the magnitude of the torsional strain, could be in principle used to design a piezoelectric sensor in WSM. The recipe to determine the torsion angle starts by experimentally measuring the differential conductance $G = dI/dV$ for a fixed and controlled value of the
external magnetic field $B_0$, as displayed in Fig.~\ref{fig4a}. As aforementioned, the plateaus in the current, and hence the peaks in the conductance, arise from the bias voltage window imposed by two consecutive Landau levels (mainly from the $\mathbf{K}_{-}$-node), i.e. $e\Delta V = eV_{n+1} - eV_{n} \sim E_{n+1}^{(-)} - E_{n}^{(-)} = \frac{\hbar v_F}{a}\sqrt{2|B_{-}|a^2/\tilde{\phi}_0}\left(\sqrt{n+1} - \sqrt{n}\right)$, where in the equation we measure energies in units of $\hbar v_F/a$ and magnetic field in units of $\tilde{\phi}_0/a^2$. Therefore, by reading the
locus of two consecutive peaks $V_{n+1} > V_{n}$ in the conductance curve (see Fig.~\ref{fig4b} for an example), it is possible to extract the value of the corresponding integer $n$ from the ratio:
\begin{eqnarray}
\frac{V_{n+1}}{V_n} \sim \frac{E_{n+1}^{(-)}}{E_{n}^{(-)}} = \sqrt{1 + \frac{1}{n}}\Longrightarrow n = \left\lfloor\frac{1}{\left(\frac{V_{n+1}}{V_n}\right)^{2}-1}\right\rceil.
\label{eq_voltage_peaks}
\end{eqnarray}
Here, the symbol $\lfloor x\rceil$ represents the nearest integer to $x$.
With the value of $n$, one can solve for the effective pseudo-magnetic field:
\begin{eqnarray}
|B_{-}| = \frac{\left(E_{n}^{(-)}/(\hbar v_F/a)\right)^2}{2 n}\frac{\tilde{\phi}_0}{a^2} \sim  \frac{\left(eV_n/(\hbar v_F/a)\right)^2}{2 n}\frac{\tilde{\phi}_0}{a^2} = \left|B_0 - B_S\right|.
\label{eq_B_extracted}
\end{eqnarray}
As a concrete example, let us take the values in Fig.~\ref{fig4b}. We have that $n\approx\left\lfloor\frac{1}{\left(7.3/5.9\right)^{2}-1}\right\rceil=\lfloor1.88\rceil=2$ and from Eq. \eqref{eq_B_extracted} we have that $B_0 - B_S=5.9^2/(2\cdot 2) \tilde{\phi}_0/a^2=8.7\tilde{\phi}_0/a^2$. Using that $B_0=22.5\tilde{\phi}_0/a^2$, this result means that $B_{S}a^2=(22.5 - 8.7)\tilde{\phi}_0  = 13.8\tilde{\phi}_0$. Using Eq.~\eqref{eq:torsionangle}, this pseudomagnetic field corresponds to $\theta\approx 10.1^\circ$, which is within 1\% of the value ($\theta=10^\circ$) used to generate the conductance curve in  Fig.~\ref{fig4b} in the first place. This simple procedure can be applied in general, and used to read off the effective torsional angle from the conductance curves.

\section{Conclusions and Summary}

We studied the electronic transport in WSM submitted to torsional strain and an external magnetic field. For this purpose we consider a cylindrical region embedded in a bulk WSM connected to two semi-infinite regions that act as reservoirs held at a different bias voltage. Using the partial wave method of scattering theory \cite{sakurai} and the analytical method developed in \cite{Munoz-2017}, we find analytical expressions for the transmission and Weyl node current components through the cylindrical region. Our analytical results predict a node-polarization effect  on the current (analogous to the valley-polarization effect observed in graphene \cite{Low_010,Settnes_016}), due to the combined effect of the local torsional strain field and the externally imposed magnetic field. This is due to the asymmetry on the spectral density arising from the two different nodes $\mathbf{K}_{\pm}$. We remark that the node filtering effect is a consequence of the symmetry breaking between the two nodes induced by the linear combination of the strain pseudomagnetic field and the $z$-component of the external magnetic field, that have different relative signs at each node: The physical magnetic field breaks time-reversal symmetry,
whereas the strain field does not. Therefore, the node-filtering effect is robust in different geometries,
and is not only a property of the cylindrical configuration chosen for the explicit analytical solution provided in our work.

As in the case of graphene \cite{Munoz-2017}, the results obtained in this paper could be used for the construction of a torsional strain sensor for WSMs. To the best of our knowledge, there are no reports in the literature about experimental measurements of torsional strain in WSM. We assume that this is due to the technical difficulties to assess small spatial displacements on the scale of a few lattice constants of the material.
However, as it was explained in detail in the previous section, our theoretical results suggest that by performing electronic conductance measurements the magnitude of such strain could in principle be inferred, opening the possibility to develop a torsional strain-meter in WSMs.

\section*{Acknowledgements}

This work was supported by Fondecyt (Chile) Grant Nos. 11160542 (R. S.-G.) and 1141146 (E. M.).

\end{document}